\documentclass[aps,showpacs,showkeys]{revtex4}

\usepackage{amssymb,amsmath}
\usepackage[pdftex]{graphicx}

\newcommand{\bd}[1]{\mbox{\boldmath$#1$}}

\newcommand{\pd}[1]{\ensuremath{[\![#1]\!]}}
\newcommand{\Pd}[1]{\ensuremath{\Bigl[\!\!\Bigl[#1\Bigr]\!\!\Bigr]}}
\newcommand{\PD}[1]{\ensuremath{\biggl[\!\!\biggl[#1\biggr]\!\!\biggr]}}

\begin{document}

\title{Collective phase description of oscillatory convection}

\author{Yoji Kawamura}
\email{ykawamura@jamstec.go.jp}
\affiliation{Institute for Research on Earth Evolution,
Japan Agency for Marine-Earth Science and Technology, Yokohama 236-0001, Japan}

\author{Hiroya Nakao}
%\email{nakao@mei.titech.ac.jp}
\affiliation{Department of Mechanical and Environmental Informatics, Tokyo Institute of Technology, Tokyo 152-8552, Japan}
%\affiliation{JST, CREST, Kyoto 606-8502, Japan}

%\author{Yoshiki Kuramoto}
%\email{kuramoto@kurims.kyoto-u.ac.jp}
%\affiliation{Research Institute for Mathematical Sciences, Kyoto University, Kyoto 606-8502, Japan}
%\affiliation{Institute for Integrated Cell-Material Sciences, Kyoto University, Kyoto 606-8501, Japan}

%\date{\today}
%\date{January 1, 2011}
%\date{October 6, 2011}
%\date{January 1, 2013}
%\date{August 18, 2013}
\date{November 20, 2013}

%% 05.45.-a Nonlinear dynamics and chaos
%% 05.45.Xt Synchronization; coupled oscillators
%% 47.15.-x Laminar flows
%% 47.15.gp Hele-Shaw flows
%% 47.55.-t Multiphase and stratified flows
%% 47.55.P- Buoyancy-driven flows; convection
%% 47.55.pb Thermal convection


%%%%% abstract
\begin{abstract}
  We formulate a theory for the collective phase description of oscillatory convection in Hele-Shaw cells.
  It enables us to describe the dynamics of the oscillatory convection
  by a single degree of freedom which we call the collective phase.
  The theory can be considered as a phase reduction method
  for limit-cycle solutions in infinite-dimensional dynamical systems,
  namely, stable time-periodic solutions to partial differential equations,
  representing the oscillatory convection.
  We derive the phase sensitivity function,
  which quantifies the phase response of the oscillatory convection
  to weak perturbations applied at each spatial point,
  and analyze the phase synchronization between two weakly coupled Hele-Shaw cells exhibiting oscillatory convection
  on the basis of the derived phase equations.
\end{abstract}

\maketitle

%%%%% lead paragraph
\begin{quotation}
{\bf
  Self-sustained oscillations and synchronization phenomena are ubiquitous in nonlinear dynamical systems,
  e.g., in biological, chemical, electrical, mechanical, neural, and optical systems.
  In many cases,
  each oscillatory unit is described by an ordinary differential equation with a stable limit-cycle orbit,
  and the phase description method~\cite{ref:winfree80,ref:kuramoto84}
  has been successfully applied to analyze weakly coupled limit-cycle oscillators.
  Synchronization of oscillatory spatiotemporal dynamics has also been observed in fluid systems
  and is potentially important in various geophysical problems.
  The oscillatory spatiotemporal dynamics is generally described
  by limit cycles of partial differential equations with an infinite-dimensional state space,
  but the phase description method has not been fully developed for such systems.
  In this paper, as the first step toward theoretical understanding of the synchronization phenomena in fluid systems,
  we formulate a phase description method for oscillatory convection in a Hele-Shaw cell.
  Using the method,
  we analyze the phase synchronization of the oscillatory convection between a pair of Hele-Shaw cells.
}
\end{quotation}

%%%%%
%\begin{quotation}
%{\bf
%  Self-sustained oscillations and synchronization phenomena are ubiquitous in nonlinear dynamical systems.
%  %%
%  Examples of such systems include biological, chemical, electrical, mechanical, neural, and optical oscillations.
%  %%
%  In many cases,
%  each oscillatory unit is described by an ordinary differential equation,
%  which has a stable limit-cycle orbit in its state space.
%  %%
%  The phase reduction method~\cite{ref:winfree80,ref:kuramoto84}
%  has been successfully applied to analyze weakly coupled limit-cycle oscillators.
%  %%
%  However, there are also abundant examples of nonlinear dynamical systems exhibiting oscillatory spatiotemporal dynamics;
%  they are generally described by partial differential equations such as reaction-diffusion equations and fluid equations.
%  %%
%  Synchronization phenomena between spatiotemporal patterns have also attracted considerable attention among researchers,
%  but a phase reduction method for partial differential equations has not been fully developed yet.
%  %%
%  In this paper, as the first step,
%  we formulate a phase reduction method for oscillatory convection in Hele-Shaw cells.
%  %%
%  The method facilitates analytical treatments of the phase synchronization between oscillatory convections.
%}
%\end{quotation}
%%%%%

%%%%% section 1
\section{Introduction} \label{sec:1}

Synchronization of oscillatory dynamics is ubiquitously observed
in real-world systems~\cite{ref:winfree80,ref:kuramoto84,ref:strogatz03}.
In the theoretical analysis,
each oscillatory unit is typically described
by a finite-dimensional ordinary differential equation possessing a stable limit-cycle orbit,
i.e., a limit-cycle oscillator.
Systems of coupled limit-cycle oscillators have been extensively investigated
and shown to exhibit various kinds of intriguing collective dynamics.
In the analysis of weakly coupled limit-cycle oscillators,
the phase description method~\cite{ref:winfree80,ref:kuramoto84}
for the limit-cycle oscillator has been successfully used.
It enables us to describe the dynamics of a limit-cycle oscillator by a single phase variable,
which facilitates detailed theoretical analysis of the synchronization dynamics of weakly coupled limit-cycle oscillators.

Spatially extended nonlinear dynamical systems can exhibit oscillatory spatiotemporal patterns,
such as the oscillatory thermal convection in fluid systems
and the spiral waves in reaction-diffusion systems~\cite{ref:cross93,ref:cross09},
and synchronization phenomena between oscillatory spatiotemporal patterns
have also attracted considerable attention recently~\cite{ref:pikovsky01,ref:boccaletti02,ref:manrubia04,ref:mikhailov06}~\footnote{
  In reaction-diffusion systems, for example,
  synchronization between two locally coupled domains of excitable media exhibiting spiral wave behavior
  using the photosensitive Belousov-Zhabotinsky reaction has been investigated in Ref.~\cite{ref:hildebrand03},
  and numerical analysis of the synchronized pulses in laterally coupled excitable fibers
  using the spatially one-dimensional FitzHugh-Nagumo equations has been performed in Ref.~\cite{ref:yanagita08}.
}.
In this case,
the oscillatory spatiotemporal pattern corresponds to a stable limit-cycle solution of a partial differential equation,
whose state space is infinite-dimensional.
Therefore,
the conventional phase reduction method for ordinary limit-cycle oscillators
can not be applied to the spatially extended systems.

In fluid systems,
several experimental and numerical studies on the synchronization of oscillatory spatiotemporal patterns have been conducted,
which are mainly motivated by the synchronization phenomena observed in geophysical fluid dynamics.
For example, experimental investigations on the synchronization of convection flows have been performed
in both periodic and chaotic regimes
in a pair of thermally coupled rotating baroclinic annulus systems~\cite{ref:read09,ref:read10}.
Numerical studies on the synchronization of spatiotemporal chaos have also been conducted
in a pair of quasi-two-dimensional channel models~\cite{ref:duane01}
and in a pair of Hele-Shaw cells~\cite{ref:bernardini04}.

In this paper, as the first step toward theoretical understanding of the synchronization phenomena in fluid systems,
we analyze oscillatory thermal convection in the Hele-Shaw cell~\cite{ref:bernardini04}.
The Hele-Shaw cell is a rectangular cavity
in which the gap between two vertical walls is much smaller than the other two spatial dimensions.
We chose this system,
because the oscillatory convection in the Hele-Shaw cell has been widely studied
and it provides a simple model for the convection flow in porous media, which is motivated by geophysical applications
(see Refs.~\cite{ref:bernardini04,ref:nield06} and also references therein).
We focus on the stable time-periodic oscillatory convection,
i.e., the limit-cycle solution of the system,
and formulate a theory for the phase description of the limit cycle.
The theory enables us to describe the dynamics of the oscillatory convection
by a single degree of freedom which we call the collective phase~\footnote{
  This theory can be considered as a phase reduction method
  for limit-cycle solutions in infinite-dimensional dynamical systems.
  Using a similar idea of phase reduction,
  we recently developed a theory for the collective phase description
  of globally coupled noisy dynamical elements exhibiting macroscopic oscillations
  in Refs.~\cite{ref:kawamura07,ref:kawamura08,ref:kawamura10,ref:kawamura11};
  the theory reduces the nonlinear Fokker-Planck equation (a partial integro-differential equation)
  to the collective phase equation (an ordinary differential equation).
  In particular, in Ref.~\cite{ref:kawamura11},
  we considered the nonlinear Fokker-Planck equation,
  which does not possess spatial translational symmetry,
  describing globally coupled noisy active rotators.
  A similar formulation for stable time-periodic solutions to reaction-diffusion systems
  has also been developed in Ref.~\cite{ref:nakao12}.
}.
On the basis of our theory,
we analyze the phase synchronization of between two weakly coupled Hele-Shaw cells exhibiting oscillatory convection.

This paper is organized as follows.
In Sec.~\ref{sec:2},
we formulate a theory for the collective phase description of oscillatory Hele-Shaw convection.
In Sec.~\ref{sec:3},
we illustrate our theory using numerical simulations of the oscillatory convection.
Concluding remarks are given in Sec.~\ref{sec:4}.

%%%%% section 2
\section{Phase description of oscillatory Hele-Shaw convection} \label{sec:2}

In this section,
we formulate a theory for the collective phase description of oscillatory Hele-Shaw convection.
The theory can be considered as an extension of our phase reduction method
for the nonlinear Fokker-Planck equation~\cite{ref:kawamura11}
%and the reaction-diffusion equation~\cite{ref:nakao12}
to an equation for oscillatory convection.

%%% subsection
\subsection{Dimensionless form of governing equations}

The dynamics of the temperature field $T(x, y, t)$ in the Hele-Shaw cell
is described by the following dimensionless form
(see Ref.~\cite{ref:bernardini04} and also references therein):
%%% eq
\begin{equation}
  \frac{\partial}{\partial t} T(x, y, t)
  = \nabla^2 T + J(\psi, T),
  \label{eq:T}
\end{equation}
where the Laplacian and Jacobian are respectively given by
%%% eq
\begin{align}
  \nabla^2 T &= \left( \frac{\partial^2}{\partial x^2} + \frac{\partial^2}{\partial y^2} \right) T, \\
  J(\psi, T) &= \frac{\partial \psi}{\partial x} \frac{\partial T}{\partial y}
  - \frac{\partial \psi}{\partial y} \frac{\partial T}{\partial x}.
\end{align}
The first and second terms on the right-hand side of Eq.~(\ref{eq:T})
represent diffusion and advection, respectively.
The stream function $\psi(x, y, t)$ is determined from the temperature field $T(x, y, t)$ as follows:
%%% eq
\begin{equation}
  \nabla^2 \psi(x, y, t) = -{\rm Ra} \frac{\partial T}{\partial x},
  \label{eq:P_T}
\end{equation}
where ${\rm Ra}$ is the Rayleigh number.
The stream function also provides the fluid velocity field, i.e.,
%%% eq
\begin{equation}
  \bd{v}(x, y, t)
  = \left( \frac{\partial \psi}{\partial y}, \, -\frac{\partial \psi}{\partial x} \right).
\end{equation}
The system is defined in the unit square: $x \in [0, 1]$ and $y \in [0, 1]$.
The boundary conditions for the temperature field $T(x, y, t)$ are given by
%%% eq
\begin{align}
    \left. \frac{\partial T(x, y, t)}{\partial x} \right|_{x = 0}
  = \left. \frac{\partial T(x, y, t)}{\partial x} \right|_{x = 1} &= 0, \label{eq:bcTx} \\
    \Bigl. T(x, y, t) \Bigr|_{y = 0}  = 1, \qquad
    \Bigl. T(x, y, t) \Bigr|_{y = 1} &= 0, \label{eq:bcTy}
\end{align}
where the temperature at the bottom ($y = 0$) is higher than at the top ($y = 1$).
The stream function $\psi(x, y, t)$ satisfies
the Dirichlet zero boundary condition on both $x$ and $y$, i.e.,
%%% eq
\begin{align}
    \Bigl. \psi(x, y, t) \Bigr|_{x = 0}
  = \Bigl. \psi(x, y, t) \Bigr|_{x = 1} &= 0, \label{eq:bcPx} \\
    \Bigl. \psi(x, y, t) \Bigr|_{y = 0}
  = \Bigl. \psi(x, y, t) \Bigr|_{y = 1} &= 0. \label{eq:bcPy}
\end{align}
Owing to this set of boundary conditions given by
Eqs.~(\ref{eq:bcTx})(\ref{eq:bcTy}) and Eqs.~(\ref{eq:bcPx})(\ref{eq:bcPy}),
the system does not possess spatial translational symmetry.

%%% subsection
\subsection{Variational components of the temperature field}

To simplify the boundary conditions in Eq.~(\ref{eq:bcTy}),
we consider the following variational component $X(x, y, t)$ of the temperature field $T(x, y, t)$:
%%% eq
\begin{equation}
  T(x, y, t) = (1 - y) + X(x, y, t).
  \label{eq:T-X}
\end{equation}
Inserting Eq.~(\ref{eq:T-X}) into Eq.~(\ref{eq:T}) and Eq.~(\ref{eq:P_T}),
we derive
%%% eq
\begin{equation}
  \frac{\partial}{\partial t} X(x, y, t)
  = \nabla^2 X + J(\psi, X) - \frac{\partial \psi}{\partial x},
  \label{eq:X}
\end{equation}
and
%%% eq
\begin{equation}
  \nabla^2 \psi(x, y, t)
  = -{\rm Ra} \frac{\partial X}{\partial x}.
  \label{eq:P_X}
\end{equation}
Applying Eq.~(\ref{eq:T-X}) to Eqs.~(\ref{eq:bcTx})(\ref{eq:bcTy}),
we obtain the following boundary conditions for $X(x, y, t)$:
%%% eq
\begin{align}
    \left. \frac{\partial X(x, y, t)}{\partial x} \right|_{x = 0}
  = \left. \frac{\partial X(x, y, t)}{\partial x} \right|_{x = 1} &= 0, \label{eq:bcXx} \\
    \Bigl. X(x, y, t) \Bigr|_{y = 0}
  = \Bigl. X(x, y, t) \Bigr|_{y = 1} &= 0. \label{eq:bcXy}
\end{align}
That is, the field $X(x, y, t)$ satisfies
the Neumann zero boundary condition on $x$
and the Dirichlet zero boundary condition on $y$.

In the derivation below,
it should be be noted that Eq.~(\ref{eq:P_X}) can also be written in the following form:
%%% eq
\begin{equation}
  \psi(x, y, t)
  = \int_0^1 dx' \int_0^1 dy' \, G(x, y, x', y')
  \frac{\partial}{\partial x'} X( x', y', t),
  \label{eq:P}
\end{equation}
where the Green's function $G(x, y, x', y')$ is the solution to
%%% eq
\begin{equation}
  \nabla^2 G(x, y, x', y')
  = -{\rm Ra} \, \delta(x - x') \, \delta(y - y'),
  \label{eq:G}
\end{equation}
under the Dirichlet zero boundary condition on both $x$ and $y$.
In the following two subsections,
we analyze the dynamical equation~(\ref{eq:X}) using Eq.~(\ref{eq:P_X}) or Eq.~(\ref{eq:P}),
under the boundary conditions given by
Eqs.~(\ref{eq:bcXx})(\ref{eq:bcXy}) and Eqs.~(\ref{eq:bcPx})(\ref{eq:bcPy}).

%%% subsection
\subsection{Time-periodic solution and its Floquet-type system}

In general, a stable time-periodic solution to Eq.~(\ref{eq:X}),
which represents oscillatory convection in the Hele-Shaw cell,
can be described by
%%% eq
\begin{equation}
  X(x, y, t) = X_0\bigl( x, y, \Theta(t) \bigr), \qquad
  \dot{\Theta}(t) = \Omega,
  \label{eq:X_X0}
\end{equation}
where $\Theta$ and $\Omega$ are the collective phase and collective frequency,
respectively~\footnote{
  The dependence of the convection in the Hele-Shaw cell on the Rayleigh number is well known,
  and the existence of stable time-periodic solutions to Eq.~(\ref{eq:X}) is also well established
  (see Ref.~\cite{ref:bernardini04} and also references therein).
}.
The time-periodic solution $X_0(x, y, \Theta)$ has
the $2\pi$-periodicity with respect to $\Theta$, i.e.,
$X_0(x, y, \Theta + 2\pi) = X_0(x, y, \Theta)$.
Inserting Eq.~(\ref{eq:X_X0}) into Eq.~(\ref{eq:X}) and Eq.~(\ref{eq:P_X}),
we find that $X_0(x, y, \Theta)$ satisfies
%%% eq
\begin{equation}
  \Omega \frac{\partial}{\partial \Theta} X_0(x, y, \Theta)
  = \nabla^2 X_0 + J(\psi_0, X_0) - \frac{\partial \psi_0}{\partial x},
  \label{eq:X0}
\end{equation}
where
%%% eq
\begin{equation}
  \nabla^2 \psi_0(x, y, \Theta) = -{\rm Ra} \frac{\partial X_0}{\partial x}.
\end{equation}
Let $u(x, y, \Theta, t)$ represent a small disturbance to the time-periodic solution $X_0(x, y, \Theta)$,
and consider a slightly perturbed solution
%%% eq
\begin{equation}
  X(x, y, t) = X_0\bigl( x, y, \Theta(t) \bigr) + u\bigl( x, y, \Theta(t), t \bigr).
\end{equation}
Equation~(\ref{eq:X}) is then linearized with respect to $u(x, y, \Theta, t)$ as follows:
%%% eq
\begin{equation}
  \frac{\partial}{\partial t} u(x, y, \Theta, t)
  = {\cal L}(x, y, \Theta) u(x, y, \Theta, t).
  \label{eq:linear}
\end{equation}
Here, the linear operator ${\cal L}(x, y, \Theta)$ is explicitly given by
%%% eq
\begin{equation}
  {\cal L}(x, y, \Theta) u(x, y, \Theta)
  = \left[ L(x, y, \Theta) - \Omega \frac{\partial}{\partial \Theta} \right] u(x, y, \Theta),
  \label{eq:calL}
\end{equation}
where
%%% eq
\begin{equation}
  L(x, y, \Theta) u(x, y, \Theta)
  = \nabla^2 u + J(\psi_0, u) + J(\psi_u, X_0) - \frac{\partial \psi_u}{\partial x}.
  \label{eq:L}
\end{equation}
Similarly to the time-periodic solution $X_0(x, y, \Theta)$,
the function $u(x, y, \Theta)$ satisfies
the Neumann zero boundary condition on $x$
and the Dirichlet zero boundary condition on $y$.
In Eq.~(\ref{eq:L}), the function $\psi_u(x, y, \Theta)$ is the solution to
%%% eq
\begin{equation}
  \nabla^2 \psi_u(x, y, \Theta) = -{\rm Ra} \frac{\partial u}{\partial x},
  \label{eq:P_u}
\end{equation}
under the Dirichlet zero boundary condition on both $x$ and $y$.
Note that ${\cal L}(x, y, \Theta)$ is periodic in time through $\Theta$,
and therefore, Eq.~(\ref{eq:linear}) is a Floquet-type system with a periodic linear operator.

The phase reduction method simplifies the dynamics of the system
by projecting it onto the phase direction along the limit cycle of the oscillatory Hele-Shaw convection.
For this purpose,
we introduce the adjoint operator ${\cal L}^\ast(x, y, \Theta)$
of the linearized operator ${\cal L}(x, y, \Theta)$
around the time-periodic solution $X_0(x, y, \Theta)$
and its zero eigenfunction $U_0^\ast(x, y, \Theta)$.
Defining the inner product of two functions as
%%% eq
\begin{equation}
  \Pd{ u^\ast(x, y, \Theta), \, u(x, y, \Theta) }
  = \frac{1}{2\pi} \int_0^{2\pi} d\Theta \int_0^1 dx \int_0^1 dy \,
  u^\ast(x, y, \Theta) u(x, y, \Theta),
  \label{eq:inner}
\end{equation}
we introduce the adjoint operator of the linear operator ${\cal L}(x, y, \Theta)$ by
%%% eq
\begin{equation}
  \Pd{ u^\ast(x, y, \Theta), \, {\cal L}(x, y, \Theta) u(x, y, \Theta) }
  = \Pd{ {\cal L}^\ast(x, y, \Theta) u^\ast(x, y, \Theta), \, u(x, y, \Theta) }.
  \label{eq:operator}
\end{equation}
By partial integration, the adjoint operator ${\cal L}^\ast(x, y, \Theta)$ is explicitly given by
%%% eq
\begin{equation}
  {\cal L}^\ast(x, y, \Theta) u^\ast(x, y, \Theta)
  = \left[ L^\ast(x, y, \Theta) + \Omega \frac{\partial}{\partial \Theta} \right] u^\ast(x, y, \Theta),
  \label{eq:calLast}
\end{equation}
where
%%% eq
\begin{equation}
  L^\ast(x, y, \Theta) u^\ast(x, y, \Theta)
  = \nabla^2 u^\ast
  + \frac{\partial}{\partial x} \left[ u^\ast \frac{\partial \psi_0}{\partial y} \right]
  - \frac{\partial}{\partial y} \left[ u^\ast \frac{\partial \psi_0}{\partial x} \right]
  + \frac{\partial}{\partial x} \Bigl[ \psi_{u,x}^\ast - \psi_{u,y}^\ast \Bigr].
  \label{eq:Last}
\end{equation}
The function $u^\ast(x, y, \Theta)$ also satisfies
the Neumann zero boundary condition on $x$
and the Dirichlet zero boundary condition on $y$.
In Eq.~(\ref{eq:Last}), the two functions,
$\psi_{u,x}^\ast(x, y, \Theta)$ and $\psi_{u,y}^\ast(x, y, \Theta)$, are the solutions to
%%% eq
\begin{align}
  \nabla^2 \psi_{u,x}^\ast(x, y, \Theta)
  &= -{\rm Ra} \frac{\partial}{\partial x}
  \left[ u^\ast \left( \frac{\partial X_0}{\partial y} - 1 \right) \right],
  \label{eq:Past_uast_x} \\
  \nabla^2 \psi_{u,y}^\ast(x, y, \Theta)
  &= -{\rm Ra} \frac{\partial}{\partial y}
  \left[ u^\ast \frac{\partial X_0}{\partial x} \right],
  \label{eq:Past_uast_y}
\end{align}
under the Dirichlet zero boundary condition on both $x$ and $y$, respectively.
Details of the derivation of the adjoint operator ${\cal L}^\ast(x, y, \Theta)$
are given in App.~\ref{sec:A}.

%%% subsection
\subsection{Floquet zero eigenfunctions}

In the calculation below,
we use the Floquet eigenfunctions associated with the zero eigenvalue, i.e.,
%%% eq
\begin{align}
  {\cal L}(x, y, \Theta) U_0(x, y, \Theta)
  = \left[ L(x, y, \Theta) - \Omega \frac{\partial}{\partial \Theta} \right] U_0(x, y, \Theta)
  &= 0, \\
  {\cal L}^\ast(x, y, \Theta) U_0^\ast(x, y, \Theta)
  = \left[ L^\ast(x, y, \Theta) + \Omega \frac{\partial}{\partial \Theta} \right] U_0^\ast(x, y, \Theta)
  &= 0. \label{eq:U0ast}
\end{align}
Note that the right zero eigenfunction $U_0(x, y, \Theta)$ can be chosen as
%%% eq
\begin{equation}
  U_0(x, y, \Theta) = \frac{\partial}{\partial \Theta} X_0(x, y, \Theta),
  \label{eq:U0}
\end{equation}
which is confirmed by differentiating Eq.~(\ref{eq:X0}) with respect to $\Theta$.
Using the inner product~(\ref{eq:inner}) with the right zero eigenfunction~(\ref{eq:U0}),
the left zero eigenfunction $U_0^\ast(x, y, \Theta)$ is normalized as
%%% eq
\begin{equation}
  \Pd{ U_0^\ast(x, y, \Theta), \, U_0(x, y, \Theta) }
  = \frac{1}{2\pi} \int_0^{2\pi} d\Theta \int_0^1 dx \int_0^1 dy \,
  U_0^\ast(x, y, \Theta) U_0(x, y, \Theta)
  = 1.
\end{equation}
Here, we note that the following equation holds
(see also Refs.~\cite{ref:kawamura11,ref:hoppensteadt97}):
%%% eq
\begin{align}
  & \frac{\partial}{\partial \Theta}
  \left[ \int_0^1 dx \int_0^1 dy \, U_0^\ast(x, y, \Theta) U_0(x, y, \Theta) \right]
  \nonumber \\
  &= \int_0^1 dx \int_0^1 dy \, \left[
    U_0^\ast(x, y, \Theta) \frac{\partial}{\partial \Theta} U_0(x, y, \Theta)
    + U_0(x, y, \Theta) \frac{\partial}{\partial \Theta} U_0^\ast(x, y, \Theta) \right]
  \nonumber \\
  &= \frac{1}{\Omega} \int_0^1 dx \int_0^1 dy \, \biggl[
    U_0^\ast(x, y, \Theta) L(x, y, \Theta) U_0(x, y, \Theta)
    - U_0(x, y, \Theta) L^\ast(x, y, \Theta) U_0^\ast(x, y, \Theta) \biggr]
  \nonumber \\
  &= 0.
\end{align}
Therefore, it turns out that the following normalization condition
is satisfied independently for each value of $\Theta$:
%%% eq
\begin{equation}
  \int_0^1 dx \int_0^1 dy \, U_0^\ast(x, y, \Theta) U_0(x, y, \Theta) = 1.
  \label{eq:normalization}
\end{equation}
In the following two subsections,
using the time-periodic solution and its Floquet zero eigenfunctions,
we formulate a theory for the collective phase description of oscillatory Hele-Shaw convection.

%%% subsection
\subsection{Oscillatory convection with weak perturbations} \label{subsec:2E}

In this subsection,
we consider a single Hele-Shaw cell exhibiting oscillatory convection with a weak perturbation to the temperature field
as described by the following equation:
%%% eq
\begin{equation}
  \frac{\partial}{\partial t} X(x, y, t)
  = \nabla^2 X + J(\psi, X) - \frac{\partial \psi}{\partial x} + \epsilon p(x, y, t).
  \label{eq:Xp}
\end{equation}
The weak perturbation is denoted by $\epsilon p(x, y, t)$.
Here, we assume that the perturbed solution is always near the limit-cycle orbit.
Using the idea of phase reduction~\cite{ref:kuramoto84},
we can derive a phase equation from the perturbed equation~(\ref{eq:Xp}).
Namely, we project the dynamics of the perturbed equation~(\ref{eq:Xp}) onto the unperturbed solution as
%%% eq
\begin{align}
  \dot{\Theta}(t)
  &= \int_0^1 dx \int_0^1 dy \, U_0^\ast(x, y, \Theta) \frac{\partial}{\partial t} X(x, y, t)
  \nonumber \\
  &\simeq \Omega + \epsilon \int_0^1 dx \int_0^1 dy \, U_0^\ast(x, y, \Theta) p(x, y, t),
\end{align}
where we approximated $X(x, y, t)$ by the unperturbed solution $X_0(x, y, \Theta)$
and used the fact that
%%% eq
\begin{equation}
  \int_0^1 dx \int_0^1 dy \, U_0^\ast(x, y, \Theta)
  \frac{\partial}{\partial t} X_0(x, y, \Theta)
  = \Omega \int_0^1 dx \int_0^1 dy \, U_0^\ast(x, y, \Theta) U_0(x, y, \Theta)
  = \Omega.
\end{equation}
Therefore, the phase equation describing the oscillatory Hele-Shaw convection with a weak perturbation
is approximately obtained in the following form:
%%% eq
\begin{equation}
  \dot{\Theta}(t)
  = \Omega + \epsilon \int_0^1 dx \int_0^1 dy \, Z(x, y, \Theta) p(x, y, t),
  \label{eq:Theta}
\end{equation}
where the {\it phase sensitivity function} is defined as
%%% eq
\begin{equation}
  Z(x, y, \Theta) = U_0^\ast(x, y, \Theta).
\end{equation}
It should be noted that Eq.~(\ref{eq:Theta}) corresponds to
a phase equation that is derived for a perturbed limit-cycle oscillator
described by a finite-dimensional dynamical system
(see Refs.~\cite{ref:winfree80,ref:kuramoto84,ref:hoppensteadt97,ref:izhikevich07,ref:ermentrout10,ref:ermentrout96,ref:brown04}).
In particular, the phase variable $\Theta(t)$ depends only on time.
However, reflecting the aspects of an infinite-dimensional dynamical system,
the phase sensitivity function $Z(x, y, \Theta)$ of the oscillatory Hele-Shaw convection
possesses infinitely many components that are continuously parameterized by the two variables,
i.e., $x$ and $y$.

Here, we describe a numerical method for obtaining the left zero eigenfunction (i.e., the phase sensitivity function).
From Eq.~(\ref{eq:U0ast}), the phase sensitivity function $Z(x, y, \Theta)$ satisfies
%%% eq
\begin{equation}
  \Omega \frac{\partial}{\partial \Theta} Z(x, y, \Theta)
  = - L^\ast(x, y, \Theta) Z(x, y, \Theta),
\end{equation}
which can be transformed into
%%% eq
\begin{equation}
  \frac{\partial}{\partial s} Z(x, y, -\Omega s)
  = L^\ast(x, y, -\Omega s) Z(x, y, -\Omega s),
  \label{eq:adjoint}
\end{equation}
where $\Theta = -\Omega s$.
To numerically calculate the eigenfunction associated with the zero eigenvalue (i.e., the phase sensitivity function),
it is convenient to evolve Eq.~(\ref{eq:adjoint}) with the normalization condition~(\ref{eq:normalization}).
Because the limit-cycle solution is linearly stable
and therefore the eigenvalues of all other eigenfunctions have negative real parts,
the functional components corresponding to non-zero eigenvalues eventually decay
and the solution converges to the phase sensitivity function with the zero eigenvalue.
For ordinary differential equations,
this method is called the adjoint method~\cite{ref:hoppensteadt97,ref:izhikevich07,ref:ermentrout10,ref:ermentrout96,ref:brown04}.
In Refs.~\cite{ref:kawamura11,ref:nakao12},
we used a similar method for partial differential equations.
%%%%%
%A relaxation method using Eq.~(\ref{eq:adjoint})
%with the normalization condition~(\ref{eq:normalization}),
%which is the so-called adjoint method
%(
%see Refs.~\cite{ref:hoppensteadt97,ref:izhikevich07,ref:ermentrout10,ref:ermentrout96,ref:brown04}
%for ordinary differential equations
%and 
%Refs.~\cite{ref:kawamura11,ref:nakao12}
%for partial differential equations
%),
%is convenient means of obtaining the phase sensitivity function.
%%%
%That is, we numerically calculates
%the eigenfunction associated with the zero eigenvalue (i.e., the phase sensitivity function)
%by evolving Eq.~(\ref{eq:adjoint}) with the normalization condition~(\ref{eq:normalization})
%until all the other eigenfunctions with negative eigenvalues decay.
%%%%%

%%% subsection
\subsection{Weakly coupled Hele-Shaw cells exhibiting oscillatory convection}

In this subsection,
we consider weakly coupled Hele-Shaw cells exhibiting oscillatory convection
described by the following equation~\cite{ref:bernardini04}:
%%% eq
\begin{equation}
  \frac{\partial}{\partial t} X_\sigma(x, y, t)
  = \nabla^2 X_\sigma + J(\psi_\sigma, X_\sigma) - \frac{\partial \psi_\sigma}{\partial x}
  + \epsilon \bigl( X_\tau - X_\sigma \bigr),
  \label{eq:Xsigma}
\end{equation}
for $(\sigma, \tau) = (1, 2)$ or $(2, 1)$,
where the stream function of each system is determined by
%%% eq
\begin{equation}
  \nabla^2 \psi_\sigma(x, y, t)
  = -{\rm Ra} \frac{\partial X_\sigma}{\partial x}.
\end{equation}
Two identical Hele-Shaw cells exhibiting oscillatory convection are mutually coupled
through corresponding temperatures at each spatial point~\footnote{
  Our formulation is applicable to any coupling form,
  e.g., asymmetric, nonlinear, spatially nonlocal, or spatially partial coupling,
  as long as the coupling intensity is sufficiently weak.
},
where the coupling parameter is denoted by $\epsilon$.
Here, we assume that unperturbed oscillatory Hele-Shaw convection is a stable time-periodic solution
and that the coupling between the two Hele-Shaw cells is sufficiently weak.
Under this assumption, as in the preceding subsection,
we can obtain a phase equation from Eq.~(\ref{eq:Xsigma}) as follows:
%%% eq
\begin{equation}
  \dot{\Theta}_\sigma(t)
  = \Omega + \epsilon \int_0^1 dx \int_0^1 dy \,
  Z(x, y, \Theta_\sigma) \Bigl( X_0(x, y, \Theta_\tau) - X_0(x, y, \Theta_\sigma) \Bigr).
  \label{eq:Theta_Winfree}
\end{equation}
Applying the averaging method~\cite{ref:kuramoto84} to Eq.~(\ref{eq:Theta_Winfree}),
we can derive the following phase equation:
%%% eq
\begin{equation}
  \dot{\Theta}_\sigma(t)
  = \Omega + \epsilon \Gamma\left( \Theta_\sigma - \Theta_\tau \right),
  \label{eq:Theta_Kuramoto}
\end{equation}
where the {\it phase coupling function} is given by
%%% eq
\begin{equation}
  \Gamma(\Theta) = \frac{1}{2\pi} \int_0^{2\pi} d\lambda \int_0^1 dx \int_0^1 dy \,
  Z(x, y, \lambda + \Theta) \Bigl( X_0(x, y, \lambda) - X_0(x, y, \lambda + \Theta) \Bigr).
  \label{eq:Gamma}
\end{equation}
The phase coupling function depends only on the phase difference,
and the type of coupling (e.g., in-phase or anti-phase)
is determined by the anti-symmetric component
of the phase coupling function~\cite{ref:kuramoto84}, i.e.,
%%% eq
\begin{equation}
  \Gamma_{\rm a}(\Theta) = \Gamma(\Theta) - \Gamma(-\Theta).
  \label{eq:Gamma_a}
\end{equation}
Finally, we note that the form of Eq.~(\ref{eq:Theta_Kuramoto}) is the same as
that of the phase equation which is derived from weakly coupled limit-cycle oscillators
described by finite-dimensional dynamical systems (see Ref.~\cite{ref:kuramoto84}).
That is, a system of oscillatory convection can be reduced to a phase oscillator,
similarly to an ordinary limit-cycle oscillator.

%%%%% section 3
\section{Numerical analysis of oscillatory Hele-Shaw convection} \label{sec:3}

In this section,
using numerical simulations of oscillatory Hele-Shaw convection,
we illustrate the theory developed in the preceding section.

%%% subsection
\subsection{Spectral transform and order parameters}

In visualizing the limit-cycle oscillation of the spatiotemporal field variable,
it is convenient to use the spectral representation of the field variable $X(x, y, t)$.
Considering the boundary conditions of $X(x, y, t)$, i.e., Eqs.~(\ref{eq:bcXx})(\ref{eq:bcXy}),
we introduce the following spectral transform:
%%% eq
\begin{equation}
  H_{jk}(t) = \int_0^1 dx \int_0^1 dy \, X(x, y, t) \cos(\pi j x) \sin(\pi k y),
\end{equation}
for $j = 0, 1, 2, \cdots$ and $k = 1, 2, \cdots$.
In visualizing the limit-cycle orbit in the infinite-dimensional state space,
we project the time-periodic solution $X_0(x, y, \Theta)$ onto the $H_{11}$-$H_{22}$ plane as
%%% eq
\begin{align}
  H_{11}(\Theta)
  &= \int_0^1 dx \int_0^1 dy \, X_0(x, y, \Theta) \cos(\pi x) \sin(\pi y), \\
  H_{22}(\Theta)
  &= \int_0^1 dx \int_0^1 dy \, X_0(x, y, \Theta) \cos(2 \pi x) \sin(2 \pi y),
\end{align}
which can be considered as a pair of order parameters
quantifying the variations in the first and second long-wavelength spectral components of the field variable.

%%% subsection
\subsection{Time-periodic solution and phase sensitivity function}

In this subsection,
we first consider a single Hele-Shaw cell exhibiting oscillatory convection
described by the partial differential equation~(\ref{eq:X}).
We use the pseudospectral method in numerical simulations,
which is a standard numerical method in computational fluid dynamics
that integrates the dynamical equation in the spectral representation
while calculating the nonlinear terms in the real-space representation
(see, e.g., Refs.~\cite{ref:fornberg98,ref:takehiro06} for pseudospectral methods).
The field variables are decomposed using
a sine expansion with $128$ modes for the Dirichlet zero boundary condition
and a cosine expansion with $128$ modes for the Neumann zero boundary condition.
The initial values were chosen
so that the system exhibits single-cellular
(i.e., one vortex) oscillatory convection~\footnote{
  For completeness, we briefly summarize the Hele-Shaw convection
  (see Ref.~\cite{ref:bernardini04} and also references therein for details).
  The critical Rayleigh number for the onset of convection is ${\rm Ra} = 4 \pi^2$.
  The convection is single-cellular convection,
  which rotates clock-wise or counter-clock-wise depending on the initial condition.
  As the Rayleigh number ${\rm Ra}$ is increased,
  the single-cellular convection exhibits the following dynamics~\cite{ref:bernardini04}:
  stationary     (  39.5 --  386.4);
  periodic       ( 386.4 --  505  );
  quasi-periodic ( 505   --  560  );
  periodic       ( 560   --  950  );
  quasi-periodic ( 950   -- 1200  );
  chaotic        (1200   --       ).
}.
The Rayleigh number was fixed to ${\rm Ra} = 480$,
giving a collective frequency of $\Omega \simeq 622$.

\begin{figure*}
  \begin{center}
    \includegraphics[width=\hsize,clip]{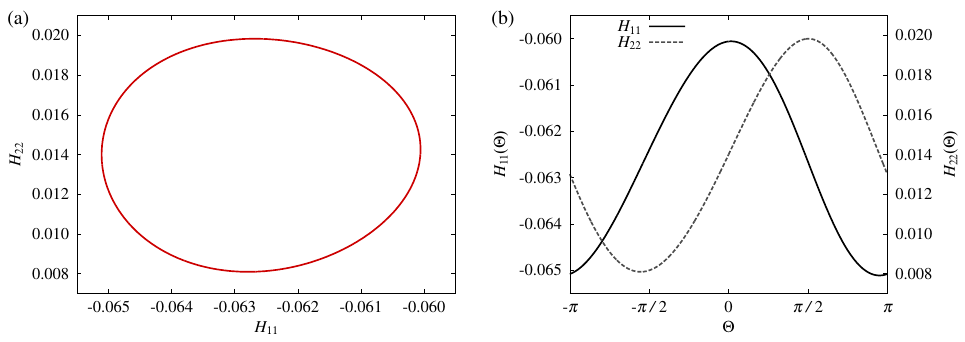}
    \caption{(Color online)
      (a) Limit-cycle orbit projected onto the $H_{11}$-$H_{22}$ plane.
      (b) Wave forms of $H_{11}(\Theta)$ and $H_{22}(\Theta)$.
      The Rayleigh number is ${\rm Ra} = 480$,
      and then the collective frequency is $\Omega \simeq 622$.
    }
    \label{fig:1}
  \end{center}
  %%%%% parameter
  %% 
  %% Rayleigh number: Ra = 480.
  %% collective frequency: \Omega = 621.666697. [622]
  %% coupling parameter: \epsilon = 0.05.
  %% 
  %% mode number for x or y: 128.
  %% mode number for \Theta: 512.
  %% 
  %% fourth-order Runge-Kutta method with integrating factor using \varDelta t = 0.000001 = 10^{-6}.
  %% fourth-order Runge-Kutta method with integrating factor using \varDelta t = 0.000100 = 10^{-4}. [DNS]
  %% 
  %%%%%
\end{figure*}

\begin{figure*}
  \begin{center}
    \includegraphics[width=\hsize,clip]{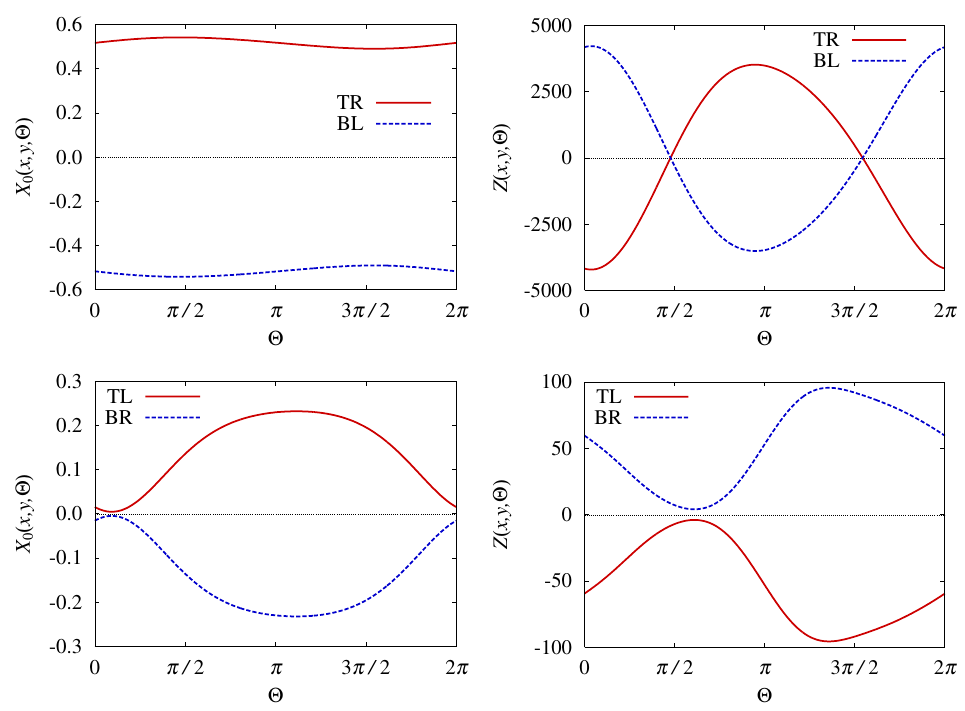}
    \caption{(Color online)
      Typical shapes of both $X_0(x, y, \Theta)$ and $Z(x, y, \Theta)$ with respect to $\Theta$
      at $(x, y) = (0.9, 0.9)$ [Top-Right (TR)],
      $(0.1, 0.1)$ [Bottom-Left (BL)],
      $(0.1, 0.9)$ [Top-Left (TL)],
      $(0.9, 0.1)$ [Bottom-Right (BR)].
      %%
      % 115.5 / 128 = 0.90234375 \simeq 0.9.
      %  12.5 / 128 = 0.09765625 \simeq 0.1.
    }
    \label{fig:3}
  \end{center}
\end{figure*}

Figure~\ref{fig:1} shows the limit-cycle orbit projected onto the $H_{11}$-$H_{22}$ plane
obtained from our numerical simulations of the dynamical equation~(\ref{eq:X}).
Snapshots of the time-periodic solution $X_0(x, y, \Theta)$ and other associated functions,
i.e., $T_0(x, y, \Theta) = (1 - y) + X_0(x, y, \Theta)$, $\psi_0(x, y, \Theta)$, $U_0(x, y, \Theta)$,
and $Z(x, y, \Theta)$, are shown in Fig.~\ref{fig:2}.
The phase sensitivity function $Z(x, y, \Theta)$
was obtained using the numerical method explained in Sec.~\ref{subsec:2E}
for the spectral representation of Eq.~(\ref{eq:adjoint}).
The typical shapes of
both the time-periodic solution $X_0(x, y, \Theta)$
and the phase sensitivity function $Z(x, y, \Theta)$ with respect to $\Theta$
are shown in Fig.~\ref{fig:3}.

Here, we note that in this case,
the phase sensitivity function $Z(x, y, \Theta)$ possesses a special property~\footnote{
  It should be noted that this property comes from the symmetry of the time-periodic solution
  under the simulation conditions performed in this paper.
  Our formulation itself is applicable to any functional form of the time-periodic solution.
}.
Namely, for each $\Theta$,
similarly to the time-periodic solution $X_0(x, y, \Theta)$,
the phase sensitivity function $Z(x, y, \Theta)$ is
anti-symmetric with respect to the center of the system:
%%% eq
\begin{align}
  X_0(-x_\delta, -y_\delta, \Theta)
  &= -X_0(x_\delta, y_\delta, \Theta), \\
  Z(-x_\delta, -y_\delta, \Theta)
  &= -Z(x_\delta, y_\delta, \Theta),
\end{align}
where $x_\delta = x - 1/2$ and $y_\delta = y - 1/2$.
Therefore, the phase sensitivity function is equal to zero at the central point,
i.e., $Z(x = 1/2, y = 1/2, \Theta) = 0$.
In addition,
the spatial integral of the phase sensitivity function also becomes zero:
%%% eq
\begin{equation}
  \int_0^1 dx \int_0^1 dy \, Z(x, y, \Theta) = 0.
\end{equation}
Namely, when the weak perturbation is spatially uniform, i.e., $p(x, y, t) = q(t)$,
the collective phase is neither advanced nor delayed by the perturbation.
It should also be noted that
the phase sensitivity function $Z(x, y, \Theta)$ is spatially localized;
the amplitudes of the phase sensitivity function $Z(x, y, \Theta)$ with respect to $\Theta$
in the top-right and bottom-left corner regions of the system
are much larger than in the other regions.

%%% subsection
\subsection{Phase synchronization between two weakly coupled Hele-Shaw cells exhibiting oscillatory convection} \label{subsec:3C}

In this subsection, as in Ref.~\cite{ref:bernardini04},
we consider two weakly coupled Hele-Shaw cells exhibiting oscillatory convection
described by the partial differential equation~(\ref{eq:Xsigma}).
Here, we assume that unperturbed oscillatory Hele-Shaw convection is described by a stable time-periodic solution
and that the coupling between the Hele-Shaw cells is sufficiently weak.
Under this assumption,
we can theoretically analyze the phase synchronization between the Hele-Shaw cells exhibiting oscillatory convection.

\begin{figure*}
  \begin{center}
    \includegraphics[width=\hsize,clip]{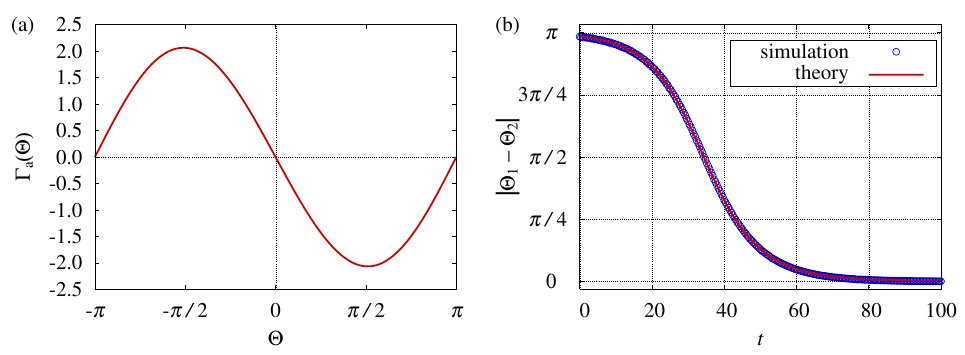}
    \caption{(Color online)
      (a) Anti-symmetric component of the phase coupling function,
      i.e., $\Gamma_{\rm a}(\Theta) = \Gamma(\Theta) - \Gamma(-\Theta)$.
      (b) Time evolution of the collective phase difference,
      i.e., $| \Theta_1 - \Theta_2 |$,
      with the coupling parameter $\epsilon = 0.05$.
    }
    \label{fig:4}
  \end{center}
\end{figure*}

The anti-symmetric component of the phase coupling function
calculated using Eqs.~(\ref{eq:Gamma})(\ref{eq:Gamma_a})
is shown in Fig.~\ref{fig:4}(a).
As can be seen,
the phase coupling function describes in-phase (attractive) coupling,
%i.e., $\Gamma_{\rm a}'(0) < 0$ and $\Gamma_{\rm a}'(\pm\pi) > 0$,
i.e., $d\Gamma_{\rm a}(\Theta)/d\Theta|_{\Theta = 0     } < 0$
and   $d\Gamma_{\rm a}(\Theta)/d\Theta|_{\Theta = \pm\pi} > 0$,
such that the two Hele-Shaw cells of oscillatory convection will become in-phase synchronized.

Figure~\ref{fig:4}(b) shows
the time evolution of the collective phase difference
$| \Theta_1 - \Theta_2 |$ between the two Hele-Shaw cells of oscillatory convection,
which started from an almost anti-phase state with the coupling parameter $\epsilon = 0.05$.
The two Hele-Shaw cells of oscillatory convection eventually became in-phase synchronized,
namely, $\Theta_1 = \Theta_2$.
Comparing our direct numerical simulation of Eq.~(\ref{eq:Xsigma})
to the theory, i.e., $\dot{\Theta} = \epsilon \Gamma_{\rm a}(\Theta)$,
we find perfect agreement between the two.

Similarly,
we can also consider phase synchronization between clock-wise convection and counter-clock-wise convection
as mentioned in App.~\ref{sec:B}.

%%%%% section 4
\section{Concluding remarks} \label{sec:4}

We developed a theory for the collective phase description of oscillatory convection in Hele-Shaw cells,
by which a system of oscillatory convection can be reduced to a phase oscillator.
On the basis of our theory,
we analyzed the phase synchronization between two weakly coupled Hele-Shaw cells exhibiting oscillatory convection.
The key component of our theory is the phase sensitivity function of the oscillatory Hele-Shaw convection,
which quantifies its phase response to weak perturbations applied at each spatial point.

The notion of {\it collective phase} used in this paper
originated from the phase of the collective oscillation
emerging from coupled individual phase oscillators
\cite{ref:kawamura07,ref:kawamura08,ref:kawamura10,ref:kawamura11}.
In this paper, as in Ref.~\cite{ref:kawamura11},
the collective phase is associated with temporal translational symmetry breaking
in partial differential equations.
In general, the phase arises not only from temporal translational symmetry breaking
but also from spatial translational symmetry breaking~\cite{ref:kuramoto84}.
In fact, the phase dynamics of spatially periodic structures,
based on spatial translational symmetry breaking,
have been extensively developed
\cite{ref:kuramoto84a,ref:kuramoto89,ref:ohta87,ref:sasa97},
and the phase dynamics approach to spatially periodic patterns
is commonly used for fluid systems
\cite{ref:pomeau79,ref:cross83,ref:cross84,ref:brand83,ref:brand84,ref:fauve87,ref:manneville90}
(see also Refs.~\cite{ref:cross93,ref:cross09}).
In addition, the so-called interface dynamics or pulse dynamics of patterns
are also essentially based on spatial translational symmetry breaking
\cite{ref:ei94,ref:ei02,ref:kilpatrick12,ref:lober12,ref:biktasheva09,ref:biktasheva10,ref:biktashev10}.
In contrast to these studies,
our formulation in this paper is based only on temporal translational symmetry breaking.
Therefore, the formulation is applicable to oscillatory Hele-Shaw convection,
although this system does not possess spatial translational symmetry owing to its boundary conditions.
It should also be noted that
the treatments of the boundary conditions for the collective phase descriptions,
including the detailed analysis of the non-trivial bilinear concomitant (see App.~\ref{sec:A}),
are newly developed in this paper for the first time,
because the nonlinear Fokker-Planck equations
studied in Refs.~\cite{ref:kawamura07,ref:kawamura08,ref:kawamura10,ref:kawamura11}
satisfy periodic boundary conditions and do not require such treatments.

We also note that the phase variable depends only on time
in Eq.~(\ref{eq:Theta}) and Eq.~(\ref{eq:Theta_Kuramoto}),
and that space-dependent phase variables can not be defined for the oscillatory Hele-Shaw convection.
For comparison, consider oscillatory reaction-diffusion systems described by
$\partial_t \bd{X}(\bd{r}, t) = \bd{F}(\bd{X}) + \hat{D} \nabla^2 \bd{X}$,
where
$\dot{\bd{X}} = \bd{F}(\bd{X})$
represents a limit-cycle oscillator located at each spatial point $\bd{r}$;
an oscillatory reaction-diffusion system can be considered as ``coupled oscillators'',
so that space-dependent phase variables can be defined,
and nonlinear phase diffusion equations,
e.g., Burgers-type equations or Kuramoto-Sivashinsky equations,
can then be derived by the conventional phase reduction method~\cite{ref:kuramoto84}.
In contrast, the oscillatory Hele-Shaw convection is described by Eq.~(\ref{eq:T}),
in which both terms on the right-hand side represent ``interactions'',
since they involve the spatial gradient.
Thus, a system of oscillatory Hele-Shaw convection can not be considered as ``coupled oscillators'',
so that space-dependent phase variables can not be defined.
In general, as mentioned in Ref.~\cite{ref:kuramoto84},
even though a fluid system exhibits oscillatory motion,
the system can not be considered as ``coupled oscillators'',
which is in sharp contrast to the oscillatory reaction-diffusion system.
The oscillatory Hele-Shaw convection is generated by the whole system,
and the oscillation is a limit-cycle solution in the infinite-dimensional state space
described genuinely by the partial differential equation.
Therefore, only the collective phase description method can be applied,
in which the collective phase is assigned to the temporal translational symmetry breaking
in the partial differential equation
and it depends only on time.
As mentioned above,
when fluid systems possess spatial translational symmetry,
conventional phase dynamics of spatially periodic structures can be developed,
in which the phase variables are space-dependent
(see, e.g., Refs.~\cite{ref:kuramoto84,ref:cross93,ref:cross09,ref:kuramoto84a,ref:kuramoto89,ref:manneville90}).
However, Hele-Shaw cells do not possess spatial translational symmetry owing to the boundary conditions,
and so the conventional phase reduction method can not be applied.

Finally, we note the broad applicability of our approach,
which is not restricted to the oscillatory Hele-Shaw convection.
If we assume that a limit-cycle solution is stable and the perturbations are sufficiently weak,
i.e., the perturbed solution is always near the limit-cycle orbit,
similarly to ordinary differential equations,
the partial differential equations can generally be reduced to phase equations by our approach.
There are abundant examples of rhythmic phenomena in nature
that can be described by partial differential equations,
such as geophysical fluid dynamics~\cite{ref:read09,ref:read10,ref:duane01,ref:bernardini04,ref:nield06},
and the phase description approach has the capability to play a central role in such areas.

%%%%% acknowledgments
\begin{acknowledgments}
  The authors are grateful to Yoshiki Kuramoto for valuable discussions.
  The first author (Y.K.) is grateful to members of both
  the Earth Evolution Modeling Research Team and
  the Nonlinear Dynamics and Its Application Research Team
  at IFREE/JAMSTEC for fruitful comments.
  The first author (Y.K.) is also grateful for financial support by
  JSPS KAKENHI Grant Number 25800222.
  The second author (H.N.) is grateful for financial support by
  JSPS KAKENHI Grant Numbers 25540108 and 22684020.
  %%
  %The second author (H.N.) is grateful for financial support by
  %JSPS KAKENHI Grant Numbers 25540108 and 22684020,
  %CREST Kokubu project of JST, and
  %FIRST Aihara project of JSPS.
  %%
  %This work was supported by JSPS KAKENHI Grant Number 25800222.
  %This work was supported by JSPS KAKENHI Grant Numbers 25800222, 25540108, 22684020.
\end{acknowledgments}

\appendix

%%%%% section A
\section{Derivation of the adjoint operator} \label{sec:A}

In this appendix,
we describe the details of the derivation of the adjoint operator ${\cal L}^\ast(x, y, \Theta)$
given in Eqs.~(\ref{eq:calLast})(\ref{eq:Last})
(see also, e.g., Refs.~\cite{ref:zwillinger98,ref:keener00} for mathematical terms).
From Eqs.~(\ref{eq:calL})(\ref{eq:L}),
the linear operator ${\cal L}(x, y, \Theta)$ is given by % the following form:
%%% eq
\begin{equation}
  {\cal L}(x, y, \Theta) u(x, y, \Theta)
  = \frac{\partial^2 u}{\partial x^2}
  + \frac{\partial^2 u}{\partial y^2}
  - \frac{\partial \psi_0}{\partial y} \frac{\partial u}{\partial x}
  + \frac{\partial \psi_0}{\partial x} \frac{\partial u}{\partial y}
  + \frac{\partial \psi_u}{\partial x} \left( \frac{\partial X_0}{\partial y} - 1 \right)
  - \frac{\partial \psi_u}{\partial y} \frac{\partial X_0}{\partial x}
  - \Omega \frac{\partial u}{\partial \Theta}.
\end{equation}
By partial integration,
each term of the inner product $\pd{u^\ast(x, y, \Theta), {\cal L}(x, y, \Theta) u(x, y, \Theta)}$
can be transformed into
%%% eq
\begin{align}
  \PD{ u^\ast, \frac{\partial^2 u}{\partial x^2} }
  &= \frac{1}{2\pi} \int_0^{2\pi} d\Theta \int_0^1 dy \,
  \left\{ \left[ u^\ast \, \frac{\partial u}{\partial x} \right]_{x=0}^{x=1}
  - \left[ \frac{\partial u^\ast}{\partial x} \, u \right]_{x=0}^{x=1} \right\}
  + \PD{ \frac{\partial^2 u^\ast}{\partial x^2}, u },
  \\
  \PD{ u^\ast, \frac{\partial^2 u}{\partial y^2} }
  &= \frac{1}{2\pi} \int_0^{2\pi} d\Theta \int_0^1 dx \,
  \left\{ \left[ u^\ast \, \frac{\partial u}{\partial y} \right]_{y=0}^{y=1}
  - \left[ \frac{\partial u^\ast}{\partial y} \, u \right]_{y=0}^{y=1} \right\}
  + \PD{ \frac{\partial^2 u^\ast}{\partial y^2}, u },
  \\
  \PD{ u^\ast, - \frac{\partial \psi_0}{\partial y} \frac{\partial u}{\partial x} }
  &= -\frac{1}{2\pi} \int_0^{2\pi} d\Theta \int_0^1 dy \,
  \left[ u^\ast \, \frac{\partial \psi_0}{\partial y} \, u \right]_{x=0}^{x=1}
  + \PD{ \frac{\partial}{\partial x} \left[ u^\ast \frac{\partial \psi_0}{\partial y} \right] , u },
  \\
  \PD{ u^\ast, \frac{\partial \psi_0}{\partial x} \frac{\partial u}{\partial y} }
  &= \frac{1}{2\pi} \int_0^{2\pi} d\Theta \int_0^1 dx \,
  \left[ u^\ast \, \frac{\partial \psi_0}{\partial x} \, u \right]_{y=0}^{y=1}
  + \PD{ -\frac{\partial}{\partial y} \left[ u^\ast \frac{\partial \psi_0}{\partial y} \right] , u },
  \\
  \PD{ u^\ast, \frac{\partial \psi_u}{\partial x} \left( \frac{\partial X_0}{\partial y} - 1 \right) }
  &= \frac{1}{2\pi} \int_0^{2\pi} d\Theta \int_0^1 dy \,
  \left[ u^\ast \, \left( \frac{\partial X_0}{\partial y} - 1 \right) \, \psi_u \right]_{x=0}^{x=1}
  + \PD{ -\frac{\partial}{\partial x} \left[ u^\ast \left( \frac{\partial X_0}{\partial y} - 1 \right) \right] , \psi_u },
  \label{eq:PD_P_u_1}
  \\
  \PD{ u^\ast, -\frac{\partial \psi_u}{\partial y} \frac{\partial X_0}{\partial x} }
  &= -\frac{1}{2\pi} \int_0^{2\pi} d\Theta \int_0^1 dx \,
  \left[ u^\ast \, \frac{\partial X_0}{\partial x} \, \psi_u \right]_{y=0}^{y=1}
  + \PD{ \frac{\partial}{\partial y} \left[ u^\ast \frac{\partial X_0}{\partial x} \right] , \psi_u },
  \label{eq:PD_P_u_2}
  \\
  \PD{ u^\ast, - \Omega \frac{\partial u}{\partial \Theta} }
  &= -\frac{\Omega}{2\pi} \int_0^1 dx \int_0^1 dy \,
  \biggl[ u^\ast \, u \biggr]_{\Theta=0}^{\Theta=2\pi}
  + \PD{ \Omega \frac{\partial u^\ast}{\partial \Theta} , u }.
\end{align}
Using the Green's function $G(x, y, x', y')$ in Eq.~(\ref{eq:G}),
the function $\psi_u(x, y, \Theta)$ given in Eq.~(\ref{eq:P_u})
can also be written in the following form:
%%% eq
\begin{equation}
  \psi_u(x, y, \Theta)
  = \int_0^1 dx' \int_0^1 dy' \, G(x, y, x', y') \frac{\partial u(x', y', \Theta)}{\partial x'}.
\end{equation}
In Eqs.~(\ref{eq:PD_P_u_1})(\ref{eq:PD_P_u_2}), we perform the following manipulations:
%%% eq
\begin{align}
  &\PD{ -\frac{\partial}{\partial x} \left[ u^\ast \left( \frac{\partial X_0}{\partial y} - 1 \right) \right] , \psi_u }
  \nonumber \\
  &= -\frac{1}{2\pi} \int_0^{2\pi} d\Theta \int_0^1 dx \int_0^1 dy \,
  \frac{\partial}{\partial x} \left[ u^\ast \left( \frac{\partial X_0}{\partial y} - 1 \right) \right] \, \psi_u
  \nonumber \\
  &= -\frac{1}{2\pi} \int_0^{2\pi} d\Theta \int_0^1 dx \int_0^1 dy \int_0^1 dx' \int_0^1 dy' \, G(x, y, x', y')
  \frac{\partial u'}{\partial x'}
  \frac{\partial}{\partial x} \left[ u^\ast \left( \frac{\partial X_0}{\partial y} - 1 \right) \right]
  \nonumber \\
  &= -\frac{1}{2\pi} \int_0^{2\pi} d\Theta \int_0^1 dx \int_0^1 dy \int_0^1 dx' \int_0^1 dy' \, G(x', y', x, y)
  \frac{\partial u}{\partial x}
  \frac{\partial}{\partial x'} \left[ {u^\ast}' \left( \frac{\partial X_0'}{\partial y'} - 1 \right) \right]
  \nonumber \\
  &= -\frac{1}{2\pi} \int_0^{2\pi} d\Theta \int_0^1 dx \int_0^1 dy \, \psi_{u, x}^\ast \frac{\partial u}{\partial x}
  \nonumber \\
  &= -\frac{1}{2\pi} \int_0^{2\pi} d\Theta \int_0^1 dy \,
  \biggl[ \psi_{u, x}^\ast \, u \biggr]_{x=0}^{x=1}
  + \PD{ \frac{\partial \psi_{u, x}^\ast}{\partial x} , u },
\end{align}
and
%%% eq
\begin{align}
  &\PD{ \frac{\partial}{\partial y} \left[ u^\ast \frac{\partial X_0}{\partial x} \right] , \psi_u }
  \nonumber \\
  &= \frac{1}{2\pi} \int_0^{2\pi} d\Theta \int_0^1 dx \int_0^1 dy \,
  \frac{\partial}{\partial y} \left[ u^\ast \frac{\partial X_0}{\partial x} \right] \, \psi_u
  \nonumber \\
  &= \frac{1}{2\pi} \int_0^{2\pi} d\Theta \int_0^1 dx \int_0^1 dy \int_0^1 dx' \int_0^1 dy' \, G(x, y, x', y')
  \frac{\partial u'}{\partial x'}
  \frac{\partial}{\partial y} \left[ u^\ast \frac{\partial X_0}{\partial x} \right]
  \nonumber \\
  &= \frac{1}{2\pi} \int_0^{2\pi} d\Theta \int_0^1 dx \int_0^1 dy \int_0^1 dx' \int_0^1 dy' \, G(x', y', x, y)
  \frac{\partial u}{\partial x}
  \frac{\partial}{\partial y'} \left[ {u^\ast}' \frac{\partial X_0'}{\partial x'} \right]
  \nonumber \\
  &= \frac{1}{2\pi} \int_0^{2\pi} d\Theta \int_0^1 dx \int_0^1 dy \, \psi_{u, y}^\ast \frac{\partial u}{\partial x}
  \nonumber \\
  &= \frac{1}{2\pi} \int_0^{2\pi} d\Theta \int_0^1 dy \,
  \biggl[ \psi_{u, y}^\ast \, u \biggr]_{x=0}^{x=1}
  + \PD{ -\frac{\partial \psi_{u, y}^\ast}{\partial x} , u },
\end{align}
where we used the following abbreviations:
%%% eq
\begin{equation}
  X_0' = X_0(x', y', \Theta), \qquad
  u' = u(x', y', \Theta), \qquad
  {u^\ast}' = u^\ast(x', y', \Theta),
\end{equation}
and defined the following functions:
%%% eq
\begin{align}
  \psi_{u, x}^\ast(x, y, \Theta)
  &= \int_0^1 dx' \int_0^1 dy' \, G(x', y', x, y)
  \frac{\partial}{\partial x'}
  \left[ {u^\ast}(x', y', \Theta) \left( \frac{\partial X_0(x', y', \Theta)}{\partial y'} - 1 \right) \right],
  \label{eq:Past_x} \\
  \psi_{u, y}^\ast(x, y, \Theta)
  &= \int_0^1 dx' \int_0^1 dy' \, G(x', y', x, y)
  \frac{\partial}{\partial y'}
  \left[ {u^\ast}(x', y', \Theta) \frac{\partial X_0(x', y', \Theta)}{\partial x'} \right].
  \label{eq:Past_y}
\end{align}
Here, we note that Eqs.~(\ref{eq:Past_uast_x})(\ref{eq:Past_uast_y})
can be derived by applying the Laplacian to Eqs.~(\ref{eq:Past_x})(\ref{eq:Past_y}), respectively.
In this way, the adjoint operator ${\cal L}^\ast(x, y, \Theta)$, defined in Eq.~(\ref{eq:operator}),
is obtained as
%%% eq
\begin{equation}
  {\cal L}^\ast(x, y, \Theta) u^\ast(x, y, \Theta)
  = \frac{\partial^2 u^\ast}{\partial x^2}
  + \frac{\partial^2 u^\ast}{\partial y^2}
  + \frac{\partial}{\partial x} \left[ u^\ast \frac{\partial \psi_0}{\partial y} \right]
  - \frac{\partial}{\partial y} \left[ u^\ast \frac{\partial \psi_0}{\partial x} \right]
  + \frac{\partial \psi_{u,x}^\ast}{\partial x}
  - \frac{\partial \psi_{u,y}^\ast}{\partial x}
  + \Omega \frac{\partial u^\ast}{\partial \Theta}.
\end{equation}
In addition, the adjoint boundary conditions are given by
%%% eq
\begin{align}
    \left. \frac{\partial u^\ast(x, y, \Theta)}{\partial x} \right|_{x = 0}
  = \left. \frac{\partial u^\ast(x, y, \Theta)}{\partial x} \right|_{x = 1} &= 0, \\
    \Bigl. u^\ast(x, y, \Theta) \Bigr|_{y = 0}
  = \Bigl. u^\ast(x, y, \Theta) \Bigr|_{y = 1} &= 0,
\end{align}
which represent the Neumann zero boundary condition on $x$
and the Dirichlet zero boundary condition on $y$.
In fact, under these adjoint boundary conditions,
the bilinear concomitant ${\cal S}[ u^\ast(x, y, \Theta), u(x, y, \Theta) ]
= \pd{ u^\ast(x, y, \Theta), {\cal L}(x, y, \Theta) u(x, y, \Theta) }
- \pd{ {\cal L}^\ast(x, y, \Theta) u^\ast(x, y, \Theta), u(x, y, \Theta) }$
becomes zero, i.e.,
%%% eq
\begin{align}
  {\cal S}\Bigl[ u^\ast(x, y, \Theta), u(x, y, \Theta) \Bigr] =
  &+ \frac{1}{2\pi} \int_0^{2\pi} d\Theta \int_0^1 dy \,
  \left[ u^\ast \, \frac{\partial u}{\partial x} \right]_{x=0}^{x=1}
  \nonumber \\
  &- \frac{1}{2\pi} \int_0^{2\pi} d\Theta \int_0^1 dy \,
  \left[ \frac{\partial u^\ast}{\partial x} \, u \right]_{x=0}^{x=1}
  \nonumber \\
  &+ \frac{1}{2\pi} \int_0^{2\pi} d\Theta \int_0^1 dx \,
  \left[ u^\ast \, \frac{\partial u}{\partial y} \right]_{y=0}^{y=1}
  \nonumber \\
  &- \frac{1}{2\pi} \int_0^{2\pi} d\Theta \int_0^1 dx \,
  \left[ \frac{\partial u^\ast}{\partial y} \, u \right]_{y=0}^{y=1}
  \nonumber \\
  &-\frac{1}{2\pi} \int_0^{2\pi} d\Theta \int_0^1 dy \,
  \left[ u^\ast \, \frac{\partial \psi_0}{\partial y} \, u \right]_{x=0}^{x=1}
  \nonumber \\
  &+ \frac{1}{2\pi} \int_0^{2\pi} d\Theta \int_0^1 dx \,
  \left[ u^\ast \, \frac{\partial \psi_0}{\partial x} \, u \right]_{y=0}^{y=1}
  \nonumber \\
  &+ \frac{1}{2\pi} \int_0^{2\pi} d\Theta \int_0^1 dy \,
  \left[ u^\ast \, \left( \frac{\partial X_0}{\partial y} - 1 \right) \, \psi_u \right]_{x=0}^{x=1}
  \nonumber \\
  &- \frac{1}{2\pi} \int_0^{2\pi} d\Theta \int_0^1 dx \,
  \left[ u^\ast \, \frac{\partial X_0}{\partial x} \, \psi_u \right]_{y=0}^{y=1}
  \nonumber \\
  &- \frac{1}{2\pi} \int_0^{2\pi} d\Theta \int_0^1 dy \,
  \biggl[ \psi_{u, x}^\ast \, u \biggr]_{x=0}^{x=1}
  \nonumber \\
  &+ \frac{1}{2\pi} \int_0^{2\pi} d\Theta \int_0^1 dy \,
  \biggl[ \psi_{u, y}^\ast \, u \biggr]_{x=0}^{x=1}
  \nonumber \\
  &- \frac{\Omega}{2\pi} \int_0^1 dx \int_0^1 dy \,
  \biggl[ u^\ast \, u \biggr]_{\Theta=0}^{\Theta=2\pi}
  = 0.
\end{align}
Each term of the bilinear concomitant ${\cal S}[ u^\ast(x, y, \Theta), u(x, y, \Theta) ]$
vanishes for the following reasons:
the first and second terms become zero
owing to the Neumann zero boundary condition on $x$ for $u$ and $u^\ast$, respectively;
the third and fourth terms, the Dirichlet zero boundary condition on $y$ for $u^\ast$ and $u$, respectively;
the fifth to tenth terms, the Dirichlet zero boundary condition on both $x$ and $y$
for $\psi_0$, $\psi_u$, $\psi_{u,x}^\ast$, and $\psi_{u,y}^\ast$;
the last term, the $2\pi$-periodicity with respect to $\Theta$ for both $u$ and $u^\ast$.

%%%%% section B
\section{Phase synchronization between clock-wise convection and counter-clock-wise convection} \label{sec:B}

\begin{figure*}
  \begin{center}
    \includegraphics[width=\hsize,clip]{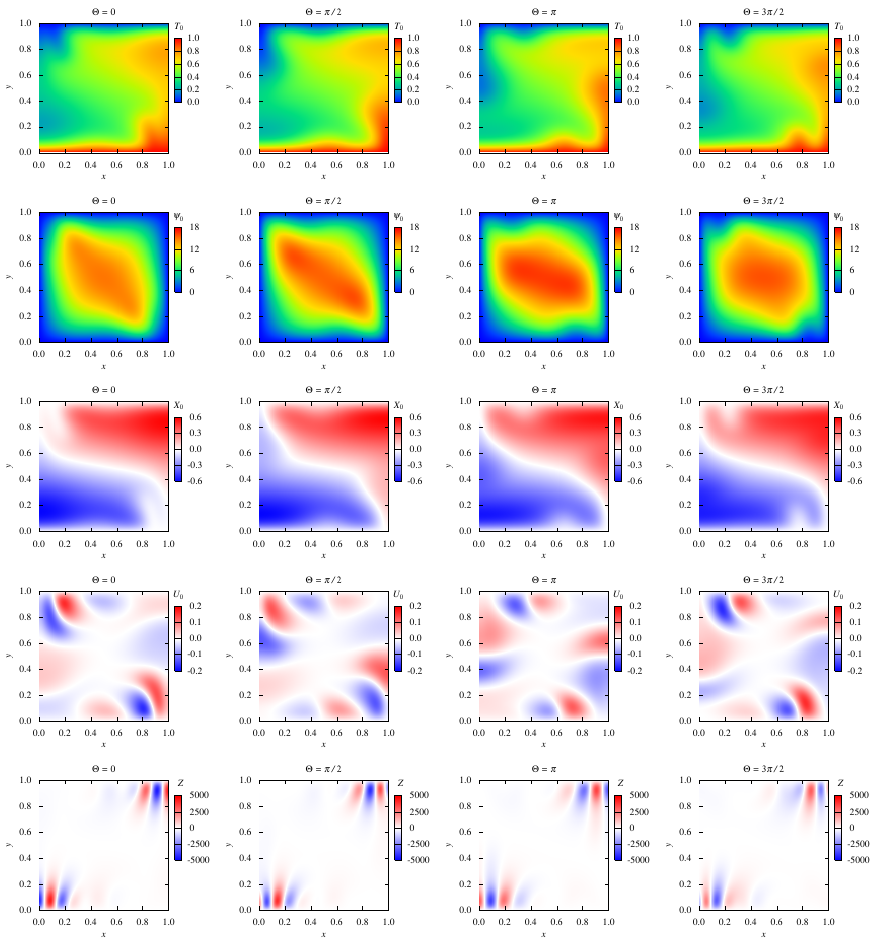}
    \caption{(Color online)
      Snapshots of
      $T_0(x, y, \Theta)$,
      $\psi_0(x, y, \Theta)$,
      $X_0(x, y, \Theta)$,
      $U_0(x, y, \Theta)$, and
      $Z(x, y, \Theta)$ for
      $\Theta = 0$, $\pi/2$, $\pi$, $3\pi/2$.
    }
    \label{fig:2}
  \end{center}
\end{figure*}

In this appendix,
we consider a supplementary problem for Sec.~\ref{subsec:3C}.
From the reflection symmetry of $x$,
the Hele-Shaw cell exhibits clock-wise convection
as well as the counter-clock-wise convection shown in Fig.~\ref{fig:2}.
Phase synchronization between the clock-wise convection and counter-clock-wise convection
can be considered as follows:
%%% eq
\begin{equation}
  \frac{\partial}{\partial t} \tilde{X}_\sigma(x, y, t)
  = \nabla^2 \tilde{X}_\sigma + J\left( \tilde{\psi}_\sigma, \tilde{X}_\sigma \right)
  - \frac{\partial \tilde{\psi}_\sigma}{\partial x}
  + \epsilon \left[ \tilde{X}_\tau(x, y, t) - \tilde{X}_\sigma(x, y, t) \right],
  \label{eq:dns_cw_ccw}
\end{equation}
for $(\sigma, \tau) = (1, 2)$ or $(2, 1)$,
where $\tilde{X}_1$ and $\tilde{\psi}_1$ correspond to the clock-wise convection,
and $\tilde{X}_2$ and $\tilde{\psi}_2$ correspond to the counter-clock-wise convection.
Here, from the reflection symmetry of $x$,
this problem is equivalent to
%%% eq
\begin{equation}
  \frac{\partial}{\partial t} X_\sigma(x, y, t)
  = \nabla^2 X_\sigma + J(\psi_\sigma, X_\sigma) - \frac{\partial \psi_\sigma}{\partial x}
  + \epsilon \Bigl[ X_\tau(1-x, y, t) - X_\sigma(x, y, t) \Bigr],
  \label{eq:cw_ccw}
\end{equation}
for $(\sigma, \tau) = (1, 2)$ or $(2, 1)$,
where both systems exhibit counter-clock-wise convection.
The only difference between Eq.~(\ref{eq:Xsigma}) and Eq.~(\ref{eq:cw_ccw}) is the $x$-dependence of $X_\tau$,
i.e., $X_\tau(x, y, t)$ in Eq.~(\ref{eq:Xsigma}) and $X_\tau(1 - x, y, t)$ in Eq.~(\ref{eq:cw_ccw}).
Therefore, a theory for the collective phase description of the system described by Eq.~(\ref{eq:cw_ccw})
can be developed in the same way~\footnote{
  Our formulation is applicable to two completely different systems of oscillatory convention,
  as long as their frequencies are near-resonant and their coupling is sufficiently weak.
  We can then determine whether the phase difference between the two systems exhibiting oscillatory convection is constant.
  However, the value of the phase difference itself is meaningful
  only when the two systems of oscillatory convection are near-identical.
  This fact is common to the conventional phase reduction method
  for ordinary limit-cycle oscillators~\cite{ref:kuramoto84}.
  From this point of view,
  phase synchronization between clock-wise convection and counter-clock-wise convection
  should be analyzed using Eq.~(\ref{eq:cw_ccw}) rather than Eq.~(\ref{eq:dns_cw_ccw}),
  as is actually done.
}.
As in Sec.~\ref{subsec:3C}, the theory indicates in-phase synchronization,
which is confirmed by direct numerical simulations of Eq.~(\ref{eq:dns_cw_ccw}) or Eq.~(\ref{eq:cw_ccw}).

%%%%% references

\clearpage

%%%%% figures

%%% fig.1
%%

%%% fig.2
%%

%%% fig.3
%%

%%% fig.4


\begin{thebibliography}{99}

%%% synchronization of limit-cycle oscillators
\bibitem{ref:winfree80}
  A.~T.~Winfree,
  {\it The Geometry of Biological Time}
  (Springer, New York, 1980; Springer, Second Edition, New York, 2001).

\bibitem{ref:kuramoto84}
  Y.~Kuramoto,
  {\it Chemical Oscillations, Waves, and Turbulence}
  (Springer, New York, 1984; Dover, New York, 2003).

\bibitem{ref:strogatz03}
  S.~H.~Strogatz,
  {\it Sync: How Order Emerges from Chaos in the Universe, Nature, and Daily Life}
  (Hyperion Books, New York, 2003).

%%% spatiotemporal pattern formation
\bibitem{ref:cross93}
  M.~C.~Cross and P.~C.~Hohenberg,
  %``Pattern formation outside of equilibrium'',
  Rev. Mod. Phys. {\bf 65}, 851 (1993).

\bibitem{ref:cross09}
  M.~C.~Cross and H.~Greenside,
  {\it Pattern Formation and Dynamics in Nonequilibrium Systems}
  (Cambridge University Press, Cambridge, 2009).

%%% synchronization of spatiotemporal chaos
\bibitem{ref:pikovsky01}
  A.~Pikovsky, M.~Rosenblum, and J.~Kurths,
  {\it Synchronization: A Universal Concept in Nonlinear Sciences}
  (Cambridge University Press, Cambridge, 2001).

\bibitem{ref:boccaletti02}
  S.~Boccaletti, J.~Kurths, G.~Osipov, D.~L.~Valladares, and C.~S.~Zhou,
  %``The synchronization of chaotic systems'',
  Phys. Rep. {\bf 366}, 1 (2002).

\bibitem{ref:manrubia04}
  S.~C.~Manrubia, A.~S.~Mikhailov, and D.~H.~Zanette,
  {\it Emergence of Dynamical Order: Synchronization Phenomena in Complex Systems}
  (World Scientific, Singapore, 2004).

\bibitem{ref:mikhailov06}
  A.~S.~Mikhailov and K.~Showalter,
  %``Control of waves, patterns and turbulence in chemical systems'',
  Phys. Rep. {\bf 425}, 79 (2006).

%%% reaction-diffusion system
\bibitem{ref:hildebrand03}
  M.~Hildebrand, J.~Cui, E.~Mihaliuk, J.~Wang, and K.~Showalter,
  %``Synchronization of spatiotemporal patterns in locally coupled excitable media'',
  Phys. Rev. E {\bf 68}, 026205 (2003).

\bibitem{ref:yanagita08}
  T.~Yanagita, H.~Suetani, and K.~Aihara,
  %``Bifurcation analysis of solitary and synchronized pulses
  %and formation of reentrant waves in laterally coupled excitable fibers'',
  Phys. Rev. E {\bf 78}, 056208 (2008).

%%% phase-separating reactive system
%\bibitem{ref:ohta05}
%  T.~Ohta and H.~Tokuda,
%  %``Dynamics of traveling waves under spatiotemporal forcing'',
%  Phys. Rev. E {\bf 72}, 046216 (2005).
%
%\bibitem{ref:tokuda07}
%  H.~Tokuda, V.~S.~Zykov, and T.~Ohta,
%  %``External forcing and feedback control of nonlinear dissipative waves'',
%  Phys. Rev. E {\bf 75}, 066203 (2007).
%
%\bibitem{ref:tonosaki10}
%  Y.~Tonosaki, T.~Ohta, and V.~S.~Zykov,
%  %``Phase description of nonlinear dissipative waves under space-time-dependent external forcing'',
%  Physica D {\bf 239}, 1718 (2010).

%%% geophysical fluid system
\bibitem{ref:read09}
  F.~J.~R.~Eccles, P.~L.~Read, A.~A.~Castrej\'on-Pita, and T.~W.~N.~Haine,
  %``Synchronization of modulated traveling baroclinic waves in a periodically forced, rotating fluid annulus'',
  Phys. Rev. E {\bf 79}, 015202(R) (2009).

\bibitem{ref:read10}
  A.~A.~Castrej\'on-Pita and P.~L.~Read,
  %``Synchronization in a pair of thermally coupled rotating baroclinic annuli:
  %Understanding atmospheric teleconnections in the laboratory'',
  Phys. Rev. Lett. {\bf 104}, 204501 (2010).

\bibitem{ref:duane01}
  G.~S.~Duane and J.~J.~Tribbia,
  %``Synchronized chaos in geophysical fluid dynamics'',
  Phys. Rev. Lett. {\bf 86}, 4298 (2001).

\bibitem{ref:bernardini04}
  A.~Bernardini, J.~Bragard, and H.~Mancini,
  %``Synchronization between two Hele-Shaw cells'',
  %Mathematical Biosciences and Engineering (MBE)
  %Math. Biosci. Eng. {\bf 1}, 339 (2004).
  Math. Biosci. Eng. {\bf 1}, 339 (2004); \\
  %%
  A.~Bernardini,
  ``Synchronization between two Hele-Shaw cells'',
  Ph.D. Thesis, University of Navarra (2005).

\bibitem{ref:nield06}
  D.~A.~Nield and A.~Bejan,
  {\it Convection in Porous Media}
  (Springer, Third Edition, New York, 2006).

%%% collective phase
\bibitem{ref:kawamura07}
  Y.~Kawamura, H.~Nakao, and Y.~Kuramoto,
  %``Noise-induced turbulence in nonlocally coupled oscillators'',
  Phys. Rev. E {\bf 75}, 036209 (2007).
  [arXiv:nlin/0702042]

\bibitem{ref:kawamura08}
  Y.~Kawamura, H.~Nakao, K.~Arai, H.~Kori, and Y.~Kuramoto,
  %``Collective phase sensitivity'',
  Phys. Rev. Lett. {\bf 101}, 024101 (2008).
  [arXiv:0807.1285]

\bibitem{ref:kawamura10}
  Y.~Kawamura, H.~Nakao, K.~Arai, H.~Kori, and Y.~Kuramoto,
  %``Phase synchronization between collective rhythms of globally coupled oscillator groups: Noisy identical case'',
  Chaos {\bf 20}, 043109 (2010).
  [arXiv:1007.4382]

\bibitem{ref:kawamura11}
  Y.~Kawamura, H.~Nakao, and Y.~Kuramoto,
  %``Collective phase description of globally coupled excitable elements'',
  Phys. Rev. E {\bf 84}, 046211 (2011).
  [arXiv:1110.0914]

%\bibitem{ref:kawamura13}
%  Y.~Kawamura,
%  %Collective phase dynamics of globally coupled oscillators: Noise-induced anti-phase synchronization,
%  submitted.

%%% Nakao paper
\bibitem{ref:nakao12}
  H.~Nakao, T.~Yanagita, and Y.~Kawamura,
  %``Phase description of stable limit-cycle solutions in reaction-diffusion systems'',
  %[2011 IUTAM Symposium on 50 Years of Chaos: Applied and Theoretical]
  Procedia IUTAM {\bf 5}, 227 (2012).

%\bibitem{ref:nakao13}
%  H.~Nakao, T.~Yanagita, and Y.~Kawamura,
%  ``Phase reduction approach to synchronization of spatiotemporal rhythms in reaction-diffusion systems'',
%  submitted.

%%% Malkin theorem
\bibitem{ref:hoppensteadt97}
  F.~C.~Hoppensteadt and E.~M.~Izhikevich,
  {\it Weakly Connected Neural Networks}
  (Springer, New York, 1997).

\bibitem{ref:izhikevich07}
  E.~M.~Izhikevich,
  {\it Dynamical Systems in Neuroscience: The Geometry of Excitability and Bursting}
  (MIT Press, Cambridge, MA, 2007).
  %www.izhikevich.com

\bibitem{ref:ermentrout10}
  G.~B.~Ermentrout and D.~H.~Terman,
  {\it Mathematical Foundations of Neuroscience}
  (Springer, New York, 2010).

\bibitem{ref:ermentrout96}
  G.~B.~Ermentrout,
  %``Type I membranes, phase resetting curves, and synchrony'',
  Neural Comput. {\bf 8}, 979 (1996).

\bibitem{ref:brown04}
  E.~Brown, J.~Moehlis, and P.~Holmes,
  %``On the phase reduction and response dynamics of neural oscillator populations'',
  Neural Comput. {\bf 16}, 673 (2004).

%%% pseudospectral method
\bibitem{ref:fornberg98}
  B.~Fornberg,
  {\it A practical Guide to Pseudospectral Methods}
  (Cambridge University Press, Cambridge, 1998).

\bibitem{ref:takehiro06}
  S.~Takehiro, M.~Odaka, K.~Ishioka, M.~Ishiwatari, and Y.-Y.~Hayashi,
  ``SPMODEL: A Series of Hierarchical Spectral Models for Geophysical Fluid Dynamics'',
  Nagare Multimedia (2006).
  http://www.nagare.or.jp/mm/2006/spmodel/ \\
  %%
  S.~Takehiro, Y.~Sasaki, Y.~Morikawa, K.~Ishioka, M.~Odaka, Y.~O.~Takahashi,
  S.~Nishizawa, K.~Nakajima, M.~Ishiwatari, Y.-Y.~Hayashi, and SPMODEL Development Group,
  ``Hierarchical Spectral Models for Geophysical Fluid Dynamics (SPMODEL)'',
  GFD Dennou Club (2011).
  http://www.gfd-dennou.org/library/spmodel/

%%% review of spatial phase
\bibitem{ref:kuramoto84a}
  Y.~Kuramoto,
  %``Phase dynamics of weakly unstable periodic structures'',
  Prog. Theor. Phys. {\bf 71}, 1182 (1984).

\bibitem{ref:kuramoto89}
  Y.~Kuramoto,
  %``On the reduction of evolution equations in extended systems: The underlying universal structure'',
  Prog. Theor. Phys. Suppl. {\bf 99}, 244 (1989).

\bibitem{ref:ohta87}
  T.~Ohta and K.~Kawasaki,
  %``Euclidean invariant phase dynamics for propagating pattern'',
  Physica D {\bf 27}, 21 (1987).

\bibitem{ref:sasa97}
  S.~Sasa,
  %``Renormalization group derivation of phase equations'',
  Physica D {\bf 108}, 45 (1997).

%%% spatial phase in fluid system
\bibitem{ref:pomeau79}
  Y.~Pomeau and P.~Manneville,
  %``Stability and fluctuations of a spatially periodic convective flow'',
  J. de Phys. Lett. {\bf 40}, 609 (1979).

\bibitem{ref:cross83}
  M.~C.~Cross,
  %``Phase dynamics of convective rolls'',
  Phys. Rev. A {\bf 27}, 490 (1983).

\bibitem{ref:cross84}
  M.~C.~Cross and A.~C.~Newell,
  %``Convection patterns in large aspect ratio systems'',
  Physica D {\bf 10}, 299 (1984).

\bibitem{ref:brand83}
  H.~R.~Brand and M.~C.~Cross,
  %``Phase dynamics for the wavy vortex state of the Taylor instability'',
  Phys. Rev. A {\bf 27}, 1237 (1983).

\bibitem{ref:brand84}
  H.~R.~Brand,
  %``Nonlinear phasedynamics for the spatially periodic states of the Taylor instability'',
  Prog. Theor. Phys. {\bf 71}, 1096 (1984).

\bibitem{ref:fauve87}
  S.~Fauve, E.~W.~Bolton, and M.~E.~Brachet,
  %``Nonlinear oscillatory convection: A quantitative phase dynamics approach'',
  Physica D {\bf 29}, 202 (1987).

\bibitem{ref:manneville90}
  P.~Manneville,
  {\it Dissipative Structures and Weak Turbulence}
  (Academic Press, New York, 1990).

%%% interface dynamics or pulse dynamics
\bibitem{ref:ei94}
  S.~Ei and T.~Ohta,
  %``Equation of motion for interacting pulses'',
  Phys. Rev. E {\bf 50}, 4672 (1994).

\bibitem{ref:ei02}
  S.~Ei, M.~Mimura, and M.~Nagayama,
  %``Pulse-pulse interaction in reaction-diffusion systems'',
  Physica D {\bf 165}, 176 (2002).

%%% traveling waves
\bibitem{ref:kilpatrick12}
  Z.~P.~Kilpatrick and G.~B.~Ermentrout, 
  %``Response of traveling waves to transient inputs in neural fields'',
  Phys. Rev. E {\bf 85}, 021910 (2012).

\bibitem{ref:lober12}
  J.~L\"ober, M.~B\"ar, and H.~Engel,
  %''Front propagation in one-dimensional spatially periodic bistable media'',
  Phys. Rev. E {\bf 86}, 066210 (2012).

%%% spiral waves
\bibitem{ref:biktasheva09}
  I.~V.~Biktasheva, D.~Barkley, V.~N.~Biktashev, G.~V.~Bordyugov, and A.~J.~Foulkes,
  %``Computation of the response functions of spiral waves in active media'',
  Phys. Rev. E {\bf 79}, 056702 (2009).

\bibitem{ref:biktasheva10}
  I.~V.~Biktasheva, A.~J.~Foulkes, D.~Barkley, and V.~N.~Biktashev,
  %``Computation of the drift velocity of spiral waves using response functions'',
  Phys. Rev. E {\bf 81}, 066202 (2010).

\bibitem{ref:biktashev10}
  V.~N.~Biktashev, D.~Barkley, and I.~V.~Biktasheva,
  %``Orbital motion of spiral waves in excitable media'',
  Phys. Rev. Lett. {\bf 104}, 058302 (2010).

%%% mathematical terms
\bibitem{ref:zwillinger98}
  D.~Zwillinger,
  {\it Handbook of Differential Equations}
  (Academic Press, Third Edition, New York, 1998).

\bibitem{ref:keener00}
  J.~P.~Keener,
  {\it Principles of Applied Mathematics: Transformation and Approximation}
  (Perseus, Second Edition, Cambridge, MA, 2000).

%%% breathing mode
%\bibitem{ref:ohta96}
%  T.~Ohta and J.~Kiyose,
%  %``Collision of domain boundaries in a reaction-diffusion system'',
%  J. Phys. Soc. Jpn. {\bf 65}, 1967 (1996).

%%% two-layer convection
%\bibitem{ref:petry03}
%  M.~Petry and F.~H.~Busse,
%  ``Theoretical study of flow coupling mechanisms in two-layer Rayleigh-Benard convection'',
%  Phys. Rev. E {\bf 68}, 016305 (2003).

%%% thermal convection of liquid metal
%\bibitem{ref:yanagisawa10a}
%  T.~Yanagisawa, Y.~Yamagishi, Y.~Hamano, Y.~Tasaka, M.~Yoshida, K.~Yano, and Y.~Takeda,
%  %``Structure of large-scale flows and their oscillation in the thermal convection of liquid gallium'',
%  Phys. Rev. E {\bf 82}, 016320 (2010).
%
%\bibitem{ref:yanagisawa10b}
%  T.~Yanagisawa, Y.~Yamagishi, Y.~Hamano, Y.~Tasaka, K.~Yano, J.~Takahashi, and Y.~Takeda,
%  %``Detailed investigation of thermal convection in a liquid metal under a horizontal magnetic field:
%  %Suppression of oscillatory flow observed by velocity profiles'',
%  Phys. Rev. E {\bf 82}, 056306 (2010).
%
%\bibitem{ref:yanagisawa11}
%  T.~Yanagisawa, Y.~Yamagishi, Y.~Hamano, Y.~Tasaka, and Y.~Takeda,
%  %``Spontaneous flow reversals in Rayleigh-Benard convection of a liquid metal'',
%  Phys. Rev. E {\bf 83}, 036307 (2011).

\end{thebibliography}
\end{document}